# Cosmic Ray Positrons at High Energies: A New Measurement

S.W. Barwick[1], J.J. Beatty[2], C.R. Bower[3], C. Chaput[4], S. Coutu[4], G. de Nolfo[2],

D. Ficenec[2], J. Knapp[5], D.M. Lowder[6], S. McKee[4], D. Müller[5], J.A. Musser[3], S.L. Nutter[4],

E. Schneider[1], S.P. Swordy[5], K.K. Tang[7], G. Tarlé[4], A.D. Tomasch[4], E. Torbet[5]

1. University of California at Irvine, Dept. of Physics, Irvine, CA 92717

2. Washington University, Dept. of Physics, St. Louis, MO 63130

3. Indiana University, Dept. of Physics, Bloomington, IN 47405

4. University of Michigan, Dept. of Physics, Ann Arbor, MI 48109

5. University of Chicago, Enrico Fermi Institute and Dept. of Physics, Chicago, IL 60637

6. University of California at Berkeley, Dept. of Physics, Berkeley, CA 94720

7. University of Utah, Dept. of Physics, Salt Lake City, UT 84112

## Abstract

We present a new measurement of the cosmic-ray positron fraction $e^+/(e^+ + e^-)$ obtained from the first balloon flight of the High Energy Antimatter Telescope (HEAT). Using a magnet spectrometer combined with a transition radiation detector, an electromagnetic calorimeter, and time-of-flight counters we have achieved a high degree of background rejection. Our results do not indicate a major contribution to the positron flux from primary sources. In particular, we see no evidence for the significant rise in the positron fraction at energies above $\sim 10$ GeV previously reported.

98.70.S, 14.80.Ly

Electrons account for one percent or less of the total cosmic ray flux, but their low mass and lack of hadronic interaction makes them subject to processes different from those governing the nuclear cosmic rays during acceleration and propagation through the Galaxy. Secondary $e^{\pm}$ are produced in about equal proportion subsequent to nuclear interactions of primary cosmic rays in interstellar space. In addition, there must be a substantial contribution of $e^-$ from primary acceleration sites since the measured positron fraction in the 1 to 10 GeV range [1,2] is less than 10%. It is not understood whether these sites are the same as those generating the nuclear cosmic rays, nor is it known why primary electrons are so much less abundant than nuclei of comparable energy. However, the flux of positrons in this energy range seems to be consistent with an entirely secondary origin [3]. At higher energies, the leaky box model of cosmic ray propagation predicts a slow decline of the secondary positron fraction while several experiments [4–8] reported a surprising rise in the positron fraction above 10 GeV. These results have motivated a variety of interpretations [9] involving either a depletion of the primary electron source at high energy or new sources of $e^{\pm}$ pairs, such as pair production near compact objects [10] or the annihilation of hypothetical dark matter particles [11]. The available data do not permit definitive conclusions among these possibilities.

The HEAT-$e^{\pm}$ instrument, shown schematically in Fig. 1, is designed to extend $e^{\pm}$ measurements to higher energies with good statistical significance and with multiple techniques for rejecting the large hadron background. It consists of a two-coil superconducting magnet with a field of $\sim$ 1T at the center, a drift tube hodoscope (DTH), a transition radiation detector (TRD), an electromagnetic shower counter (EC) and a time-of-flight system (TOF). HEAT was flown on May 3–5, 1994 from Ft. Sumner, New Mexico, and collected data for 29.5 hours at float altitudes of 3.8–7.4 g/cm$^2$ of residual atmosphere.

The DTH measures the rigidity, R, and the sign of the particle charge. It contains 479 drift tubes of 2.5 cm diameter filled with $CO_2$:hexane (96:4), 18 layers in the bending plane and 8 layers in the non-bending plane. Timing signals are measured and converted into "impact parameters" (the closest distance between the wire and the particle trajectory).

For impact parameters r > 0.20 cm, the single tube resolution is typically $\sigma \simeq 75\mu$m. Rigidities are determined by finding the best-fit track through the known magnetic field. The maximum detectable rigidity (MDR) distribution has a mean of 170 GV permitting reliable measurements up to 50 GeV.

The TOF system is designed to provide effective up/down discrimination and good charge resolution to separate singly charged particles from He nuclei. It determines the direction of particle travel and the particle velocity between the TOF scintillator and the EC with a timing accuracy of $\sigma = 0.75$ ns. The TOF scintillator also determines the charge of each particle with a resolution $\sigma = 0.11$ e. The probability of a He nucleus being mis-identified by the TOF as a singly charged particle is below $10^{-3}$ and the probability that it might survive the subsequent $e^{\pm}$ selections is negligible.

The TRD is comprised of six radiator/multiwire proportional chamber (MWPC) pairs. The radiators consist of plastic fiber blankets [12] and the MWPCs are filled with a Xe:CH$_4$ (70:30) mixture. Transition radiation signals are expected for $e^{\pm}$ but not for hadrons. Total charge signals are read from cathode strips, and clusters of charge are identified from a fast readout of sense wire signals in 25 ns time slices. A likelihood technique is used to analyze the total charge signals, and the time slice data are interpreted using a neural network technique. The response functions for different particles in each TRD layer are obtained in accelerator calibrations and confirmed with the flight data using an iterative procedure. The neural network is trained with a sample of $e^{\pm}$ candidates, and its output value ranging from 0–1 identifies $e^{\pm}$ events clustering near unity.

The EC identifies electrons as particles that deposit a large energy with a pulse height profile consistent with the development of an electromagnetic shower. We use 10 layers of Pb and plastic scintillator with 0.9 radiation lengths (r.l.) of Pb per layer. To increase the dynamic range, each photomultiplier tube (PMT) signal is processed through two pulse height analysis chains with different gains, and care is taken that the PMT signals stay well below saturation. The signals are converted into an estimate of the primary energy, E, using a covariance analysis based on the results of GEANT [13] simulations and of accelerator calibrations. The resulting energy resolution is $\sigma \simeq 7\%$–$11\%$, varying slightly with zenith angle. It is roughly independent of energy due to fluctuations in the amount of energy exiting the back of the EC. Two parameters, $\chi^2_{\text{EC}}$ for the shower fit and $X_{\text{start}}$, the shower start depth, are used to distinguish electromagnetic showers from hadronic background.

Data are telemetered to a ground station for recording and on-line display. The event trigger is formed from the TOF and EC signals. Normally the EC threshold is set to exclude non-interacting particles, but a prescaler is also used to accept a fraction ($\sim 2\%$) of penetrating protons. A subsequent slow trigger requires a minimum signal in the DTH before the event is accepted. The acceptance for particles satisfying the trigger is 320 cm$^2$sr.

To obtain a sample of clean e$^\pm$ events the data are subjected to two categories of selections shown in Table I. The first extracts events with a single downward-going particle having a unit charge and a well resolved momentum. The requirement of a single and consistent track in both the TRD and the DTH is particularly important in rejecting events in which interactions occur within the instrument.

The second category selects for e$^\pm$ by combining the hadron rejection afforded by the TRD electron likelihood analysis, the TRD time slice data, and the EC shower shape and starting depth. The proton rejection factor at an electron efficiency of 90% is 200 for the TRD and $\gtrsim 100$ for the EC. The final selection of events is based on the agreement between E and momentum p. The ratio E/p is expected to peak sharply at unity in the case of electrons, and to exhibit a much broader distribution peaking at an E/p value of less than unity in the case of interacting hadrons. The E/p ratio provides additional hadron rejection and an estimate of the residual background in the selected data. Fig. 2 shows a histogram of E/p before and after the e$^\pm$ selections are applied. The solid curve represents all data having $E_t > 4.5$ GeV *before* the e$^\pm$ selections are applied. ($E_t$ is the e$^\pm$ energy corrected to the top of the atmosphere; see below.) For positive E/p, the distribution is heavily contaminated by interacting protons and has a different shape than that for negative E/p data, which contains mostly electrons. The hatched region shows the data *after* the e$^\pm$ selections are applied. The dashed curve is the result of a Monte-Carlo calculation of the response of the EC and DTH detectors to electrons, which takes into account bremsstrahlung by e$^\pm$ in the instrument and overlying material. Bremsstrahlung photons deposit their energy in the EC but result in a lower momentum in the DTH, leading to a tail for large $|E/p|$. The agreement between the shapes of the expected and observed E/p distributions is quite good, indicating that the instrumental response of the EC and DTH detectors are well understood.

The remaining background in the final data set is small and occurs primarily at low $|E/p|$ values (see Fig. 2) as would be expected for a residual hadronic background. The data used in the determination of the positron fraction are required to satisfy the condition $0.7 < |E/p| < 3.0$ (*cross*-hatched). Because the energy and momentum measurements have been verified to be charge symmetric, this selection does not introduce a bias into the measured positron fraction. A worst case estimate of the proton contamination in the region of accepted positron candidates would indicate a background contribution of 10% to the positron flux. However, taking the shape of the distribution of interacting hadrons properly into account, we conclude that the remaining background in the positron sample is only 1%. This background is subtracted to obtain the final result.

Including the hadron rejection obtained with the EC trigger, an overall background rejection of better than $10^5$ is achieved by the data selections described above. We emphasize that these selections are not biased by charge sign dependent effects. The background distribution shown in Fig. 2 does not reflect the full rejection power of the instrument since it does not include events rejected by the trigger or events with $E_t <$ 4.5 GeV. The total electron efficiency obtained with this analysis is $\sim$ 30%. Roughly 50% of all e$^\pm$ events are rejected by the requirement of track consistency in the TRD and DTH. The remaining inefficiency, reflected in the distributions shown in Fig. 2, results from the e$^\pm$ selections.

In Table II we show the results of this analysis. The energy of each event, $E_t$, is obtained after correcting the measured energy, E, by $\sim$ 5–10% to account for radiative losses in the atmosphere. A Monte Carlo program [13] is used to determine the fraction of observed e$^\pm$ which are generated by primary cosmic rays in the atmosphere. The atmospheric correction calculation has been verified by measurements of the rate of growth of the secondary e$^+$ and

e⁻ intensity as a function of atmospheric depth. The calculated background is subtracted from the observed number of events to produce the corrected counts. The calculation includes geomagnetic cutoff rigidities and penumbral effects [14] and is normalized to published spectra of primary protons and electrons [15–18]. Uncertainties in these spectra lead to a systematic error in the atmospheric correction of about 30% corresponding to a systematic shift of approximately ±0.01 in the reported positron fractions for all energy intervals. Table II reports statistical errors corresponding to 68.3% Bayesian confidence intervals.

The positron fraction vs. energy is plotted in Fig. 3 along with a number of previous measurements. Also shown are calculations from the leaky box model [3]. The dark matter annihilation model of ref. [11] for WIMP masses of 90 and 120 GeV is superimposed upon the leaky box curve. The rise in the positron fraction seen previously is not indicated by our data. This may be due to the fact that none of the previous experiments employed all of the hadron rejection techniques available with the HEAT-e$^\pm$ experiment. The energy dependence of the positron fraction reported here appears to be consistent with the prediction of a standard leaky box model which assumes that pion decay is the dominant source of positrons in the Galactic cosmic radiation. While numerically our positron fraction appears to be slightly higher than the prediction, we must bear in mind that the prediction is subject to normalization errors and uncertainties in the choice of model parameters. Further analysis and measurements with improved statistics and extension to higher energies are required to resolve these issues and to investigate the dark matter annihilation scenario.

We gratefully acknowledge the efforts of personnel from the National Scientific Balloon Facility and we acknowledge contributions from D. Bonasera, P. Dowkontt, E. Drag, D. Ellithorpe, M. Gebhard, W. Johnson, P. Koehn, D. Kouba, D. Levin, A. Manson, T. Miller, S. Oser, J. Pitts, A. Richards, J. Robbins, G. Simberger, M. Solarz, G. Spiczak and S. Swan. This work was supported by NASA grants NAGW-2487, NAGW-2990, NAGW-1035, NAGW-1642, NAGW-1995, and NAGW-2000 and by financial assistance from our universities. J.K. acknowledges support from the Alexander von Humboldt Foundation.

FIGURES

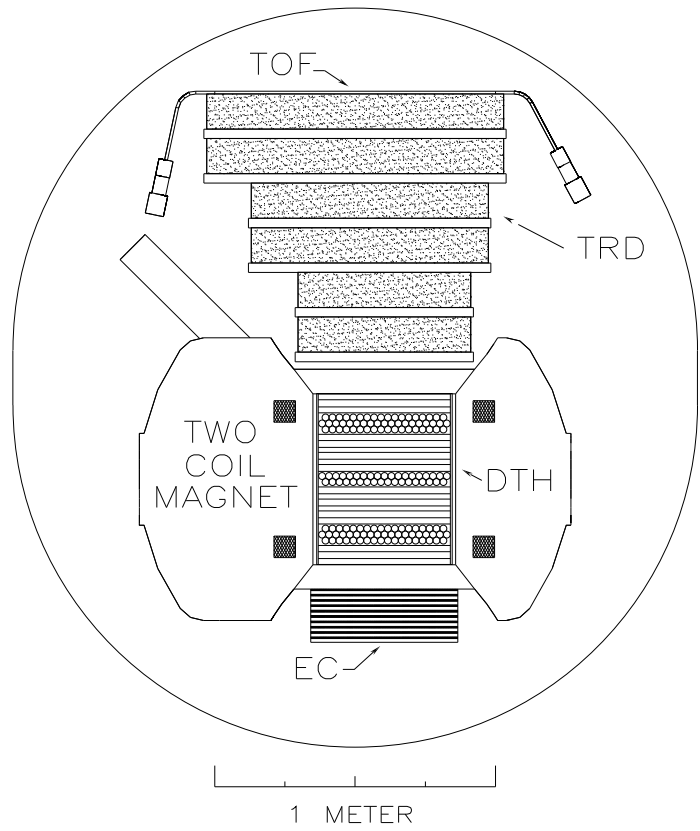

FIG. 1. Schematic Cross Section of the HEAT-e$^{\pm}$ spectrometer.

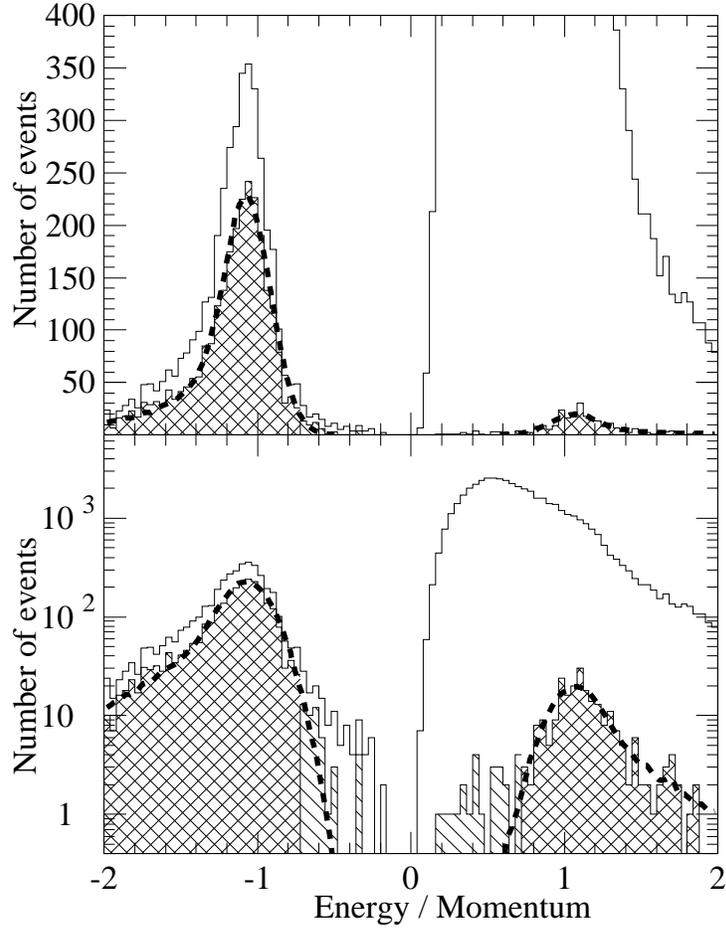

FIG. 2. The ratio of energy from the EC measurement to particle momentum from the magnetic spectrometer plotted on both linear and log scales for clarity. The solid curve is for data with group 1 selections applied, and with E > 4.5 GeV. The dashed curve is for $e^{\pm}$ from a Monte-Carlo calculation. The measured data are shown hatched with all selection criteria except $|E/p| > 0.7$. The *cross*-hatched region is the final selected data sample.

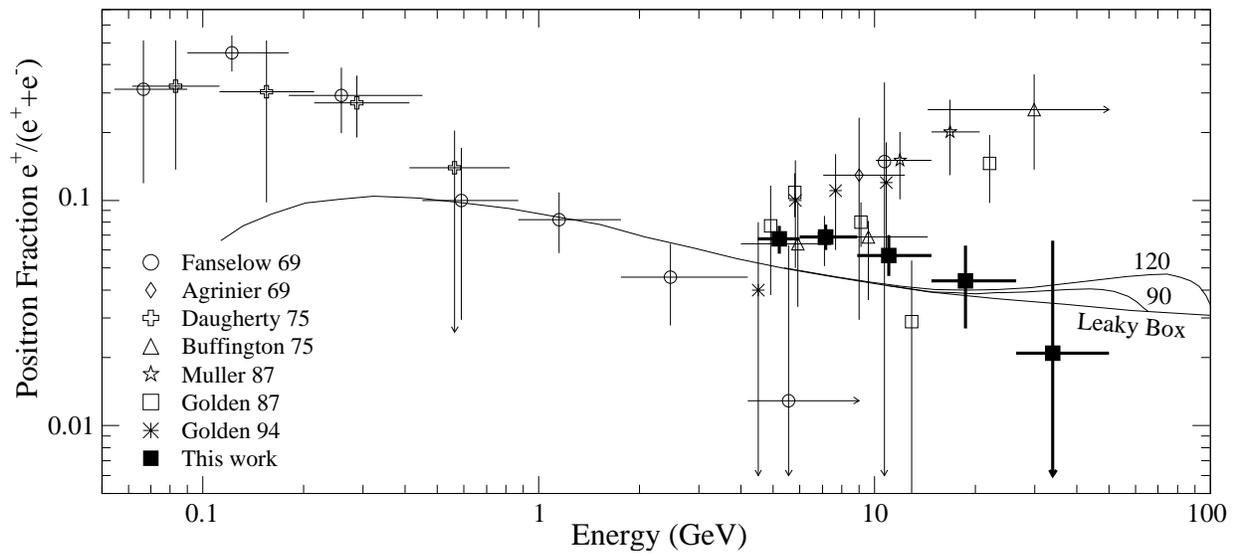

FIG. 3. A comparison of the positron fraction $e^+/(e^+ + e^-)$ as measured in this experiment along with previous measurements and theoretical models. Errors in the positron fraction are statistical.

TABLES

TABLE I. Data Selections. $\text{Int}_{\text{DTH}}$ is the intercept of the DTH track, extrapolated into the TRD, and $\text{Int}_{\text{TRD}}$ is the intercept of the TRD track. The DTH track $\chi^2$ is based on the deviation of the measured and calculated track points. MDR is the maximum detectable rigidity of the spectrometer for a given magnetic field integral, track point measurement error, and number of points used in the track fit. The TRD maximum likelihood (M.L.), the time slice neural net output, and the EC shower shape are described in the text.

| Selection Description | Selection Range |
| --- | --- |
| **Group 1** | |
| TRD, DTH track match | $|\text{Int}_{\text{DTH}} - \text{Int}_{\text{TRD}}| < 24$ cm |
| Charge = 1 | 0.77 e < Z < 1.5 e |
| Velocity = c | $0.5 < \beta < 2.0$ |
| DTH track $\chi^2$ | $\chi^2 < 10.0$ |
| DTH rigidity error | MDR/|R| > 4 |
| **Group 2** | |
| TRD $e^\pm$ M.L. | $\log(\text{M.L.}) > 2$ |
| # TRD chambers hit | $N_{\text{TRD}} = 6$ |
| TRD time slice | Neural Net output > 0.5 |
| EC shower shape | $\chi^2_{\text{EC}} < 1.8$ |
| EC shower start | $X_{\text{start}} < 0.8$ r.l. |
| energy, momentum selection | E > 3 GeV, |p| > 2.5 GeV/c |
| |E/p| | 0.7 < |E/p| < 3.0 |

TABLE II. Compilation of e$^{\pm}$ Results.

| Energy      | $< E_t >$ | e$^+$ |       | e$^-$ |        | e$^+$/(e$^+$ + e$^-$) |
|-------------|-----------|-------|-------|-------|--------|------------------------|
| $E_t$(GeV)  | (GeV)     | meas. | corr. | meas. | corr.  | corrected              |
| 4.5–6.0     | 5.22      | 113   | 75.6  | 1091  | 1046.  | $0.067^{+0.010}_{-0.009}$ |
| 6.0–8.9     | 7.17      | 107   | 75.9  | 1068  | 1030.  | $0.069^{+0.009}_{-0.009}$ |
| 8.9–14.8    | 11.0      | 51    | 33.7  | 582   | 562.0  | $0.057^{+0.013}_{-0.011}$ |
| 14.8–26.5   | 18.7      | 19    | 10.3  | 232   | 223.5  | $0.044^{+0.019}_{-0.017}$ |
| 26.5–50.0   | 34.       | 3     | 0.87  | 42    | 40.1   | $0.021^{+0.045}_{-0.021}$ |